\documentstyle[preprint,aps,eqsecnum]{revtex}
\tightenlines
\def\m@thcombine#1#2{%
  \setbox0=\hbox{$#1$}
  \setbox1=\hbox{$#2$}
  \ifdim\wd0>\wd1
    \setbox0=\hbox to\wd1{\hss\box0\hss}
  \else
    \setbox1=\hbox to\wd0{\hss\box1\hss}
  \fi
  \mathop{\vcenter{
    \offinterlineskip\box0\box1}}}
\def\lesim{\m@thcombine<\sim}
\def\gesim{\m@thcombine>\sim}
\begin{document}

\draft

\title{CAN INSTANTONS SATURATE THE LARGE MASS OF THE $\eta'$ MESON IN THE
       CHIRAL LIMIT? }

\author{V. Gogohia }

\address{HAS, CRIP, RMKI, Depart. Theor. Phys., P.O.B. 49,
          H-1525, Budapest 114, Hungary \\
         email address: gogohia@rmki.kfki.hu }

\maketitle

\begin{abstract}
 Using the trace anomaly and low energy relations, as well as the
Witten-Veneziano formula for the mass of the $\eta'$ meson, the chiral
topology of the QCD nonperturbative instanton vacuum has been numerically
evaluated. Our formalism makes it possible to express the topological susceptibility and the mass of the $\eta'$ meson as a functions of the instanton number 
density in the chiral limit. We have explicitly shown that the topological     
susceptibility in this case is approximately one fours and one half of its     
phenomenological value at instanton number densities 
in the chiral limit, $0.5 \ fm^{-4}$ and $1.0 \ fm^{-4}$, respectively. Thus   
the instanton contributions substantially
underestimate the mass of the $\eta'$ meson which still remains large in the   
chiral limit. With the help of the above-mentioned Witten-Veneziano formula,   
we derived an absolute lower bounds for the pion decay constant and the mass the $\eta'$ meson in the chiral limit.   
\end{abstract}

\pacs{PACS numbers: 11.15.Tk, 12.38.Lg }

\vfill
\eject

\section{Introduction }

It is well known that one of the most important aspects of the famous $U(1)$ problem [1,2] is the large mass of the $\eta'$ meson. It does not vanish in the chiral limit, so the $\eta'$ meson is not the Nambu-Goldstone (NG) boson. In Ref. [3] (see also Ref. [4]) by using the large $N_c$ limit technique the expression for the mass of the $\eta'$ meson was derived, namely

\begin{equation}
m^2_{\eta'} = {2 N_f \over F^2_{\pi}} \chi_t + \Delta ,
\end{equation}
where $\Delta = 2 m^2_{K} - m^2_{\eta}$, $N_f$ is the number of light quarks and $F_{\pi}$ is the pion decay constant. However, the important quantity which enters this formula is the topological density operator (topological susceptibility), $\chi_t$ (for definition see section 3). In the chiral limit it
is screened that is why it is defined for Yang-Mills (YM) fields,
i.e., for pure gluodynamics ($N_f=0$). It is one of the main characteristics
of the QCD nonperturbative vacuum where it measures the fluctuations of the    
topological charge. 

 The precise validity of the Witten-Veneziano (WV) formula (1.1) is, of course,
not completely clear because of its origin. Nevertheless, let us regard it as  
exact for simplicity (in any case we have nothing better than Eq. (1.1). However, there are phenomenological reasons [5,6] as well as some lattice indications
[7] to believe that QCD is close to $SU(\infty)$).       
Using now experimental values of all physical quantities
entering this formula, one obtains that the phenomenological ("experimental")
value of the topological susceptibility is

\begin{equation}
\chi_t^{phen} = 0.001058 \ GeV^4 = (180.36 \ MeV )^4 = 0.1377 \ GeV/fm^3.
\end{equation}
In the chiral limit $\Delta = 0$ since $K^{\pm}$ and $\eta$ particles are
NG excitations.
It is worth noting further that neither the mass of the $\eta'$ meson nor the pion decay constant in the chiral limit cannot exeed their experimental values.
So the WV formula (1.1) provides an absolute lower bounds for the pion decay constant  and the mass of the $\eta'$ meson in the chiral limit, namely

\begin{equation}
854 \leq m^0_{\eta'} \leq 957.77 \ (MeV),
\end{equation}
and

\begin{equation}
83.2 \leq F^0_{\pi} \leq 93.3 \ (MeV),
\end{equation}
respectively. They
should be compared with their experimental values (upper bounds in the previous
expressions). Let us note that the chiral perturbation theory value of the pion
decay constant in the chiral limit, $ F^0_{\pi} = (88.3 \pm 1.1) \ MeV$ [8],  
obviously satisfies these bounds, Eq. (1.4). Recent lattice result [9] (see    
also brief review [7]) for    
the mass of the $\eta'$ meson in the continuum chiral limit is $m^0_{\eta'} =
863(86) \ MeV$. It obviously satisfies our bounds, Eq. (1.3), as it should be.

One can conclude in that the mass of the $\eta'$ meson remains large even in   
the chiral limit, which is real problem indeed.
Thus the large mass of the $\eta'$ meson in the chiral limit is due to the phenomenological value of the topological susceptibility.
In other words, it is clear that through the topological susceptibility (i.e.,
via the WV formula (1.1)) the large mass of the $\eta'$ meson even in the
chiral
limit is determined by the topological properties of the QCD ground state, its nonperturbative vacuum. It has a very rich dynamical and topological structure
[10-12]. It is a very complicated medium and its dynamical and topological complexity means that its structure can be organized at various levels (quantum, classical). It can contain many different components and ingredients which may
contribute to the truly nonperturbative vacuum energy density (VED). It is well
known that VED in general is badly divergent [13], however the truly nonperturbative VED is finite, automatically negative and it has no imaginary part (stable
vacuum). For gauge-invariant definition and concrete examples see recent papers
[14]. Precisely this quantity is one of the main characteristics of the QCD
ground state and precisely it is related to the nonperturbative gluon condensate via the trace anomaly relation [15] (see section 2) as well as to the above-mentioned topological susceptibility via the low energy "theorem" (relation)    
derived by Novikov, Schifman, Vanshtein and Zakharov (NSVZ) a long time ego    
[16] and rederived
quite recently by Halperin and Zhitnitsky (HZ) [17] (see section 3).
Let us remind that the truly nonperturbative VED is nothing else but the bag
constant apart from the sign, by definition [13,14,18]. It is much more general
quantity than the string tension because it is relevant for light quarks as    
well.

 Many models of the QCD vacuum involve some extra classical color field
configurations (such as randomly oriented domains of constant
color magnetic fields, background gauge fields, averaged over
spin and color, stochastic colored background fields, etc.) and
ingredients such as color-magnetic and Abelian-projected
monopoles (see Refs. [10,11,19] and references therein).  The
relevance of center vortices for QCD vacuum by both lattice [20]
and analytical methods [21] was recently investigated as well.

However, the most elaborated classical models are the random and
interacting instanton liquid models (RILM and IILM) of the QCD
vacuum. They are based on the existence of the topologically
nontrivial, instanton-type fluctuations of gluon fields there,
which are nonperturbative, weak coupling limit solutions to the
classical equations of motion in Euclidean space [22] (and
references therein). That is instantons may be $qualitatively$
responsible for the $\eta'$ mass for the first time has been
pointed out by 't Hooft [23]. The WV formula (1.1) clearly shows
that the topological susceptibility due to instantons should be
nonzero. Here we would like to investigate this problem
$quantitatively$ as well.
In this paper we precisely address the problem whether instantons
can $quantitatively$ saturate the large mass of the $\eta'$ meson
in the chiral limit or not. The reason is that the instanton-type fluctuations 
in the QCD vacuum are totally suppressed in the chiral limit and they are      
again        
restored due to dynamical breakdown of chiral symmetry [22] (and references    
therein). Thus an important question immediately arises,
namely is this restoration in terms of instanton number densities             
$quantitatively$ sufficient for the above-mentioned purpose or not.

In order to make the problem more transparent, let us elaborate on this      
point im more detail. There exists a "standard", rather quilitative estimate   
of the topological susceptibility in terms of the instanton number density     
valid for dilute system of weakly interacting charges, namely 
$\chi_t \simeq (N / V)$ (see, for example a brief review in Ref. [7]).
For its phenomenological value, $n = (N/V) = 1 \ fm^{-4}$, one obtains
$\chi_t \simeq 0.0015 \ GeV^4$, which rather roughtly reproduces the           
phenomenological value of the topological susceptibility shown in Eq. (1.2). 
However, in the $chiral$ limit it is impossible to use the phenomenological value of the instanton number density. As was mentioned above, the 
instanton-type fluctuations are strongly suppressed in the $chiral$ 
limit and it is obvious that the "restored" value due to dynamical chiral symmetry breakdown (which is not due to instantons only) cannot be equal to its phenomenological value. Apparently it is substantially less. In any case one should
look for a possible enhancement of the right hand side of the above-mentioned  
rough estimate in
the chiral limit through the appropriate numerical coefficient in order to saturate the phenomenological value of the topological susceptibility in the left  
hand side of this estimate.\footnote{It has been already well known to NSVZ    
[16] that the right hand side of such type of relation should be in general somehow enhanced. However, their mechanism for a possible enhancement cannot be   
applied to the chiral limit case since it is completely opposite to the word   
with no 
light quarks (or, what is just the same, to the word were all quarks are very  
heavy). Moreover, in the same paper they argued that the gluon condensate in   
this limit by factor of two is less as compared to its phenomenological value  
(see discussion in sect. V below).} 
The right hand side of this estimate in the chiral limit      
stands for $n_0=(N/V)_0$ (subscript "0" means the chiral limit) while the left 
hand side is the phenomenological topological susceptibility indeed (the quenched topological susceptibility in the chiral limit). Otherwise it is zero since 
topological charge is totally screened in the chiral limit. 
Thus the real problem arises how to find a correct $quantitative$ relation     
between the topological susceptibility and the instanton number density in the 
chiral limit.      
It becomes perfectly clear that more sophisticated methods are needed for this 
purpose. The above-discussed rough estimate even on account 
of a some well-justified value for $(N/V)_0$ (which is very doubtful)          
definately fails.

One of the main purposes in this paper is to develop a     
well-justified formalism in order to establish precisely the above-mentioned   
correct quantitative relation. For the reader's convenience in
sections 2, 3 and 4 we have modified an analytical formalism (along with       
lines of NSVZ paper [16]) which has been
developed earlier [24]. It is based on using the above-mentioned
trace anomaly relation [15] (section 2), low energy "theorem" (relation)
[16,17] (section 3) and WV formula (1.1) in the chiral limit
(section 4). It allows one to directly calculate the topological
susceptibility and the mass of the $\eta'$ meson in the chiral
limit as a functions of the truly nonperturbative YM VED 
due to instantons or equivalently as a functions of the instanton number density in the chiral limit. In section 5 we present its
estimates in the chiral limit as it follows from phenomenology and lattice     
approach. The numerical results are shown in Tables I, II and III (section 6). 
Section 7 is devoted to discussion and conclusions.

\section{ The trace anomaly relation }

The truly nonperturbative VED is important in its own right as one of the
main characteristics of the QCD nonperturbative vacuum.
Furthermore it assists in estimating such an important
phenomenological parameter as the gluon condensate, introduced in
the QCD sum rules approach to resonance physics [25]. The famous
trace anomaly relation [15] in the general case (nonzero current
quark masses $m_f^0$) is

\begin{equation}
\Theta_{\mu\mu} = {\beta(\alpha_s) \over 4 \alpha_s}
G^a_{\mu\nu} G^a_{\mu\nu}
+ \sum_f m_f^0 \overline q_f q_f.
\end{equation}
where $\Theta_{\mu\mu}$ is the trace of the energy-momentum tensor
and $G^a_{\mu\nu}$ being the gluon field strength tensor while
$\alpha_s = g^2/4 \pi$.
Sandwiching Eq. (2.1) between vacuum states and on account of the
obvious relation $\langle{0} | \Theta_{\mu\mu} | {0}\rangle = 4
\epsilon_t$, one obtains

\begin{equation}
\epsilon_t = {1 \over 4}
\langle{0} | {\beta (\alpha_s)  \over 4 \alpha_s} G^a_{\mu\nu} G^a_{\mu\nu} |
 {0}\rangle
+ {1 \over 4} \sum_f m^0_f \langle{0} | \overline q_f q_f | {0}\rangle,
\end{equation}
where $\epsilon_t$ is the sum of all possible independent, truly               
nonperturbative contributions to VED (the total VED) and $\langle{0} |
 \overline q_f q_f | {0}\rangle$ is the quark condensate.
Since in what follows we want to saturate the total VED by
instantons only, i.e., to put  $\epsilon_t = \epsilon_I + ...$,
then it is legitimate to use weak coupling limit solution to the
$\beta$-function

\begin{equation}
\beta(\alpha_s) = - b { \alpha^2_s \over 2 \pi} + O(\alpha^3_s), \qquad
 b = 11 - {2 \over 3} N_f,
\end{equation}
where recalling $N_f$ is the number of light flavors. Then from Eq. (2.2)
in the chiral limit ($m^0_f=0$), one obtains

\begin{equation}
\epsilon_I = - {b \over 4} \times {1 \over 8}
\langle{0} | {\alpha_s \over \pi } G^a_{\mu\nu} G^a_{\mu\nu} | {0}\rangle.
\end{equation}
In general, of course, it is impossible to use the
above-mentioned weak coupling limit solution to the
$\beta$-function and one needs to introduce a new quantity, namely
the gluon condensate in the strong coupling limit [24].
Let us also emphasize that in the chiral limit the truly nonperturbative VED
is nothing else but the gluon condensate apart from the overall numerical      
factor (see Eqs. (2.2) and (2.4)).

\section{ The topological susceptibility}

One of the main characteristics of the QCD nonperturbative vacuum is
 the topological density operator (topological susceptibility)
in gluodynamics ($N_f=0$) [3]

\begin{equation}
\chi_t = \lim_{q \rightarrow 0} i \int d^4x\, e^{iqx} {1 \over N_c^2} \langle{0} | T \Bigl\{ {q(x) q(0)} \Bigr\} | {0} \rangle,
\end{equation}
where $q(x)$ is the topological charge density, defined as
$q(x) = (\alpha_s / 4 \pi) F (x) \tilde{F} (x) = (\alpha_s / 4 \pi)
F^a_{\mu \nu} (x) \tilde {F}^a_{\mu \nu}(x)$
and $\tilde {F}^a_{\mu \nu}(x) = (1 / 2)
\epsilon^{\mu \nu \rho \sigma} F^a_{\rho \sigma} (x)$
is the dual gluon field strength tensor while $N_c$ is the number of different 
colors. In the definition of the topological
susceptibility (3.1) it is assumed that the corresponding regularization
and subtraction of all types of the perturbative contributions have been       
already done in order Eq. (3.1) to stand for the renormalized, finite and      
the truly nonperturbative topological susceptibility (see Refs. [3,16,17,26]).
The anomaly equation in the WV notations is

\begin{equation}
\partial_{\mu} J_5^{\mu} =2 N_f (2 / N_c) (\alpha_s / 4 \pi) F \tilde{F}.
\end{equation}

The topological susceptibility
can be related to the nonperturbative gluon condensate via the
low energy "theorem" in gluodynamics proposed by NSVZ [16] (by using the       
dominance of self-dual fields hypothesis in the YM vacuum) as follows:

\begin{equation}
\lim_{q \rightarrow 0} i \int d^4x\, e^{iqx} \langle{0} | T \Bigl\{ {\alpha_s \over 8 \pi} G \tilde{G} (x)
{\alpha_s \over 8 \pi} G \tilde {G}(0) \Bigr\} | {0}
\rangle = {1 \over 2 b }\langle{0} | {\alpha_s \over \pi }
 G^a_{\mu\nu} G^a_{\mu\nu} | {0}\rangle.
\end{equation}  
Quite recently it was discussed by HZ in Ref. [17] (see also references        
therein) who noticed that it is not precisely a Ward identity, but rather 
is a relation between the corresponding correlation functions, indeed. That is 
why in what follows we call Eq. (3.3) as low energy relation or NSVZ-HZ        
relation. The anomaly equation in the NSVZ-HZ notations is

\begin{equation}
\partial_{\mu} J_5^{\mu} = N_f (\alpha_s / 4 \pi) G \tilde{G},
\end{equation}
with $N_f=3$. Thus to get the topological susceptibility in the WV form from this relation, it is necessary to make a replacement in its left hand side as follows: $G \tilde{G} \longrightarrow (2 / N_c)  F \tilde{F}$. Then the WV topological susceptibility (3.1) finally becomes    

\begin{equation}
\chi_t =  {1 \over 2 b }\langle{0} | {\alpha_s \over \pi }
 G^a_{\mu\nu} G^a_{\mu\nu} | {0}\rangle = 
- \Bigr( {4 \over b} \Bigr)^2 \epsilon_{YM},
\end{equation}
where $b=11$. The second equality comes from Eq. (2.4) by denoting the truly nonperturbative VED due to instantons at $N_f=0$ as $\epsilon_{YM}$. 
The significance of this formula is that it gives the topological
susceptibility as a function of the truly nonperturbative VED for pure gluodynamics, $\epsilon_{YM}$.

\section{ The $U(1)$ problem}

 The topological susceptibility (3.1) assists
in the resolution of the above-mentioned $U(1)$
problem [1,2]
via the WV formula for the mass of the $\eta'$ meson (1.1).
 Within our notations it is expressed as follows
$f^2_{\eta'} m^2_{\eta'} = 4 N_f \chi_t$,
where $f_{\eta'}$ is the $\eta'$ residue defined as
$\langle {0}| \sum_{q=u,d,s} \overline q \gamma_{\mu} \gamma_5 q | {\eta'}
\rangle = i \sqrt{N_f} f_{\eta'} p_{\mu}$ and $\langle {0}| N_f {\alpha_s \over 4 \pi}
 G \tilde{G} | {\eta'} \rangle = (N_c \sqrt{N_f} / 2) f_{\eta'} m^2_{\eta'}$.
 Using the normalization relation $f_{\eta'} = \sqrt{2} F^0_{\pi}$, one
 finally obtains

\begin{equation}
F^2_{\pi}m^2_{\eta'} = 2 N_f \chi_t.
\end{equation}
Eq. (3.5) then leads to

\begin{equation}
m^2_{\eta'} = - 2 N_f \Bigl( {4 \over b F_{\pi} } \Bigr)^2 \epsilon_{YM},
\end{equation}
which expresses the mass of the $\eta'$ meson as a function of the truly nonperturbative YM VED, in particularly due to instantons in this case.
In previous expressions we omit for simplicity the
superscript "0" in the pion decay constant as well as in $m^2_{\eta'}$. 

In order to directly apply this formalism to
the instanton liquid model we need the realistic estimate of the corresponding
truly nonperturbative chiral VED in this model.

\section{ The truly nonperturbative VED due to instantons}

  The contribution to the truly nonperturbative VED at the classical level from
the instanton-type nonperturbative fluctuations of gluon fields can be estimated as follows. Let us consider the trace anomaly relation (2.4) in the chiral limit again. The phenomenological analysis of QCD sum rules for the gluon condensate implies

\begin{equation}
G_2 \equiv \langle G^2 \rangle \equiv
 \langle {\alpha_s \over \pi} G^2 \rangle \equiv
\langle{0}|{\alpha_s \over
\pi}G^a_{\mu\nu}G^a_{\mu\nu}|{0}\rangle \simeq  0.012 \ GeV^4,
\end{equation}
which can be changed within a factor of two [25].
From the phenomenological  estimate (5.1), one easily can calculate

\begin{equation}
{1 \over 8} \langle{0} | (\alpha_s / \pi) G^a_{\mu\nu} G^a_{\mu\nu}
 | {0}\rangle \simeq  0.0015 \ GeV^4 \simeq  1.0 \ fm^{-4}.
\end{equation}
  Having in mind this and assuming that the gluon condensate in the weak
coupling limit is determined by the instanton-type fluctuations only, Shuryak
[22] (see also references therein) has concluded in that the "average separation"
between instantons was $\simeq 1.0 \ fm$, so the corresponding density of
the instanton-type fluctuations should be $\simeq 1.0 \ fm^{-4}$. Let us note
that the second parameter of the instanton liquid model of the QCD vacuum,
the instanton size $\rho_0 \simeq 1/3$, was chosen to reproduce standard (also 
(as gluon condensate)
 phenomenologically estimated from QCD sum rules [25]) value
of the quark condensate. This contribution to VED via the
trace anomaly relation (2.1-2.2) vanishes in the chiral limit.  However, due to
all reasonable estimates of light quark masses, numerically its contribution
is at $20\%$ and thus comparable with the systematic error in the determination
of the gluon condensate itself [25,27].

Using the above-mentioned estimate (5.2),
from Eq. (2.4) for dilute ensemble, one finally obtains

\begin{equation}
\epsilon_I = - {b \over 4} \times n_0,                                         
\end{equation}
where, let us remind, we denote the instanton number density in the chiral limit as $n_0 =(N/V)_0$ while in the general (nonchiral case) it is $n=(N/V)$.
It is well known that density of instanton-type fluctuations is strongly
suppressed in the chiral limit and is again restored bacause of
dynamical breakdown of chiral symmetry (see Ref. [22] and
references therein). In any case it can not be large in the
chiral limit, so the functional dependence of VED on the instanton number      
density in the chiral limit, established in Eq. (5.3) due to dilute gas
approximation, seems to be justified in this case. The only
problem is its numerical value which, in general of course, can not be equal   
to its phenomenological value, $1.0 \ fm^{-4}$. 
Let us emphasize that the instanton contribution to the truly
nonperturbative VED was $not \ calculated$ independently but was
$postulated$ via the trace anomaly relation (2.2) using the
phenomenological value of the gluon condensate (5.1) as well as
weak coupling limit solution to the $\beta$-function (2.3). The
significance of this expression is that it allows one to estimate
instanton's contribution to the truly nonperturbative VED as a
function of the instanton number density in the chiral limit, which can be     
estimated either from
phenomenology or taken from lattice simulations (see below).

 In Ref. [16] it has been argued that the gluon condensate in the chiral limit 
is approximately two times less than the above-mentioned its phenomenological
(empirical) value (5.1), i. e.
   $\langle G^2 \rangle_{ch} \simeq 0.5 \langle G^2 \rangle_{phen}$.
This means that in this case the instanton number density in the chiral limit  
is $n_0 \simeq 0.5 \ fm^{-4}$. However, it has been already pointed out [28]   
that QCD sum rules substantially
underestimate the value of the gluon condensate.
The most recent phenomenological calculation of the gluon condensate is
given by Narison in Ref. [29], where a brief review of many previous           
calculations is also presented.
His analysis leads to the update average value as

\begin{equation}
\langle{0}|{\alpha_s \over
\pi}G^a_{\mu\nu}G^a_{\mu\nu}|{0}\rangle = (0.0226 \pm 0.0029) \ GeV^4.
\end{equation}
This means that instanton density is approximately two time bigger than it was
estimated by Shuryak for instanton liquid model [22], but in the chiral limit
we are again left with $n_0 \simeq 1.0 \ fm^{-4}$. In Ref. [30] the dimensionless instanton density
has been evaluated through the QCD asymptotic scale parameter, $\Lambda_{QCD}$.
At a value $\Lambda_{QCD} = 280 \ MeV$  estimated in the $\overline{MS}$ scheme
for $N_f=3$ (compatible with DIS and other data), one gets the instanton density about $1.0 \ fm^{-4}$. Thus in the chiral limit we are left with about half  
of this number again. Unfortunately, the conversion to physical units is       
rather ambiguous in the pure gauge theory. It strongly depends on 
the renormalization scheme chosen for calculation. 
   
In lattice QCD situation with instanton density and their sizes
is also ambigious. In quenched ($N_f=0$) lattice QCD by using the
so-called "cooling" method the role of the instanton-type
fluctuations in the QCD vacuum was investigated [31]. In
particular, it was found that the instanton density should be
$n=(1+ \delta) \ fm^{-4}$, where $\delta \simeq 0.3-0.6$
depending on cooling steps. Moreover, by studying the topological
content of the vacuum of $SU(2)$ pure gauge theory using a method
of RG mapping [32], it is concluded that the average radius of an
instanton is about $0.2 \ fm$, at a density of about $2 \
fm^{-4}$. However, in Ref. [33] the topological content of the
$SU(3)$ vacuum was studied using the same method as for $SU(2)$
gauge theory earlier and was obtained a fair agreement with
Shuryak's phenomenologically estimated numbers for the instanton
liquid model. At the same time, in Refs. [34-36] considerably
larger values were reported and advocated. Thus at this stage it
is rather difficult to choose some well-justified numerical value for
the instanton number density in the chiral limit. If $n \simeq 2 \ fm^{-4}$,   
then we are left
with half of this value in the chiral limit, i.e., $n_0 \simeq 1 \ fm^{-4}$,   
but if $n \simeq 1 \ fm^{-4}$, we will be left with half of this, i.e.,  $n_0 \simeq 0.5 \ fm^{-4}$. 
That is why we will perform all our calculations on account of the two 
different values for the instanton number
densities in the chiral limit, $n_0 = 0.5 \ fm^{-4}, \ 1.0 \ fm^{-4}$.

In conclusion, let us note that for densities $n > 2 \
fm^{-4}$ (which means $n_0 > 1 \ fm^{-4}$) the
applicability of the dilute gas approximation becomes doubtful.               


\section{Numerical results}

Lattice approach shows that in
instanton calculus much more convenient to express all results as a functions  
of the instanton number density, $n$ [37]. Thus let us also express all our    
results 
as a functions of the instanton number density in the chiral limit, $n_0$. In  
this case, the truly nonperturbative VED is given in
Eq. (5.3) and its numerical results are shown in Table I. It explicitly
depends on $N_f$ and instanton number density $n_0$. The quark part is one     
order of magnitude less than gluon part and is of opposite sign.               
Combining Eqs. (2.4) and (5.3), one obtains

\begin{equation}
G_2 \equiv  \langle {\alpha_s \over \pi} G^2 \rangle  =
- {32 \over b} \epsilon_t \simeq - {32 \over b} \epsilon_I = 8 \times n_0,
\end{equation}
i.e., the gluon condensate in the weak coupling limit does not
explicitly depend on $N_f$ (for numerical results see Table I as
well). As it was mentioned above, precisely this gluon condensate
was introduced a long time ago [25]. This unphysical situation
takes place because in instanton calculus [22] there is no other
way to estimate the truly nonperturbative VED than the trace
anomaly relation (2.4) which becomes finally Eq. (6.1) as it was
described above. In this case it is preferable to have the $N_f$
dependent VED than the gluon condensate since the former 
mainly characterisizes the detail structure of the nonperturbative vacuum while
the latter is one of its average (global) characteristics, indeed.

The numerical values of the topological susceptibility due to instanton number 
density in the chiral limit are shown in Table II. They are determinded by the 
following relation

\begin{equation}
\chi_t = {4 \over b} \times n_0, \quad  b=11,
\end{equation}
which, obviously follows from Eqs. (3.5) and (5.3).  

The numerical values of the $\eta'$ meson in the chiral limit due to instanton 
number density in the chiral limit are shown in Table III. They are determinded
by the following relation

\begin{equation}
m^2_{\eta'} = {8 N_f \over b} {1 \over F_{\pi}^2}  \times n_0, \quad b=11,
\quad N_f=3,
\end{equation}
which follows from Eqs. (4.2) and again (5.3). 
As in previous expression (4.2) we omit for simplicity the
superscript "0" in the pion decay constant as well as in
$m^2_{\eta'}$.

\section{Discussion and conclusions}

In summary, using the trace anomaly relation (2.4), The NSVZ-HZ
low energy relation (3.3) and WV formula for the mass of the $\eta'$ meson (4.1), the chiral topology of the QCD nonperturbative instanton vacuum   
has been numerically evaluated.                                                
The NSVZ-HZ low energy relation (3.3) (which is valid in the $chiral$
limit as well (with the corresponding value of the gluon condensate in this limit)) in combination with the trace anomaly relation in the chiral limit (2.4)  
provides precisely the well-justified formalism to find the above-mentioned correct $quantitative$ relation between the topological susceptibility and the instanton nunber density in the chiral limit, Eq. (6.2). It turned out that the correct coefficient in this relation is even substantially less than one ($4 / b$
with $b=11$), so there is no way for instantons to saturate the phenomenological value of the topological susceptibility in the chiral limit.             
      
There are, of course, very strong theoretical reasons to think that the 
low energy relation (3.3) is very important. Let us remind the redear
that behind this low energy theorem-relation is a beatiful physical idea [16], 
namely that the self-dual gluon fields may be dominant in the YM vacuum. This
idea is not only beatiful but it is very powerful as well. As it is shown here 
(see also Ref. [24]), it allows to relate many important quantities such as
the truly nonperturbative VED, gluon condensate, topological susceptibility,   
density numbers of different types of excitations and fluctuations (not only 
those of instantons) of the gluon field configurations, etc. to each other in  
a well-justified way, i.e., by providing well-justified coefficients between   
the above-mentioned quantities of the same dimension. At the same time, the    
rough  
estimate discussed in the Introduction  is rather artificial. In fact, its     
right hand side has been adjusted by Shuryak by using the pnenomenological value of the gluon condensate which precisely via the relation (3.3) is related to 
the topological susceptibility. Thus precisely it justifies (and not vice      
versa) the existence 
of the above-mentioned estimate. At the same time, it is, of
course, much more general since it can be applied to the chiral limit while the
estimate certainly fails in this limit as it has been explained in the Introduction.

The topological susceptibility due to 
instanton number densities in the chiral limit (Table II) is substantially
less than its phenomenological value shown in Eq. (1.2).
Therefore they (instantons) cannot account for the large mass of
the $\eta'$ meson in the chiral limit alone (compare bounds (1.3)
with numerical results presented in Table III). This means that the
truly nonperturbative VED (the bag constant, apart from the sign,
by definition)  due to instantons (see Table I) is an order of
magnitude less than it is required to saturate the large mass of
the $\eta'$ meson in the chiral limit as well as the
phenomenological value of the topological susceptibility.                      
In turn, this means that larger (than it can be
reasonably estimated from phenomenology or lattice simulations) value
of the instanton number density in the chiral limit is required for this       
purpose. This is impossible since in the chiral limit it cannot
be large as it was underlined a few times in the main body of the text.

It is well known from recent [32,37] and old [38] lattice calculations that    
instantons by themselves do not confine quarks contributing only no more than
ten percent into the phenomenological value of the string
tension. It has been already known for a long time that 
instantons give rise to the constant (not linear) potential at
large distances in continuum theory [39]. Obviously, if instantons are not     
able to saturate the large mass of the $\eta'$ meson in the chiral limit (see  
Table III and Eq. (1.2)),
this means that they have nothing to do with its experimental
value. Thus one can conclude in that instantons provide resolution of the large
mass of the $\eta'$ meson, indeed, but $only \ quilitatively$. Their $quantitative$ contribution is not clearly sufficient. 

If instantons by themselves cannot saturate the phenomenological value of the
topological susceptibility (and consequently the large mass of
the $\eta'$ meson) in continuum theory, then a question
immediately arises how they can saturate these quantities in
lattice calculations. There are no doubts left that all lattice
data which nicely saturate the phenomenological value of the
topological susceptibility (see for example, a brief recent review in Ref. [7])
clearly indicate the existence and importance of such types of the
nonperturbative excitations of gluon field configurations in the
QCD vacuum which $cannot$ be treated as instantons. The presense of other      
(than instantons) nonperturbative fluctuations and excitations in the QCD      
true ground state has been pointed out and discussed in Ref. [40] as well.

Concluding, let us emphasize once more that instantons by themselves do not    
confine quarks and cannot saturate the phenomenological value of the          
topological susceptibility either. In order to explain confinement and         
saturate the large mass of the $\eta'$ meson in the chiral limit,              
we need completely $different \ from \ instantons$ types of the                
nonperturbative excitations of gluon field configurations in the QCD true      
vacuum. These types of vacuum excitations should have at least an order of     
magnitude larger amplitudes than those of instantons can provide at all and
they should be closely related to the excitations responsible for quark        
confinement [24,41].                                                           

Obvious simplicity of analytical
calculations of the above-discussed quantities in comparison with
obvious complexity of their calculations by lattice method (see,
for example, Refs. [7,42]) should be also mentioned.

 The author is grateful to M. Faber and A. Ivanov for many interesting remarks 
and useful discussion in the Institute of Nuclear Physics of the Thechnical   
University in Wien. It is also a        
pleasure to thank Gy. Kluge for many useful remarks and help.

\vfill

\eject

\vfill

\eject

\begin{table}
\caption{The truly nonperturbative VED and gluon condensate due to instanton number densities in the chiral limit}
\begin{center}
\renewcommand{\arraystretch}{1.5}
\begin{tabular}{lrrrrrr}
$n_0 \ (fm^{-4})$ &$\epsilon_I \ (GeV^4)$ & $N_f=0$ & $N_f=1$ & $N_f=2$ & $N_f=3$& $G_2 \ (GeV^4)$ \\
\hline
0.5 &  & $-0.002085$ & $-0.00196$ & $-0.001835$ & $-0.00171$ & 0.006 \\
1.0 &  & $-0.00417$ & $-0.00392$ & $-0.00367$ & $-0.00342$ & 0.012 \\
\end{tabular}
\renewcommand{\arraystretch}{1.0}
\end{center}
\end{table}

\begin{table}
\caption{The topological susceptibility due to instanton number densities in   
the chiral limit }
\begin{center}
\renewcommand{\arraystretch}{1.5}
\begin{tabular}{lrr}
$n_0 \ (fm^{-4})$  & $\chi_t \ (GeV^4)$ & $\chi_t^{1/4} \ (MeV)$ \\
\hline
0.5 & 0.000275 & 128 \\
1.0 & 0.00055 & 153 \\
\end{tabular}
\renewcommand{\arraystretch}{1.0}
\end{center}
\end{table}

\begin{table}
\caption{The mass of the $\eta'$ meson ($MeV$ units) due to instanton number  
 densities in the chiral limit }
\begin{center}
\renewcommand{\arraystretch}{1.5}
\begin{tabular}{lrr}
$n_0 \ (fm^{-4})$ & $m_{\eta'}^0 \ (F^0_{\pi}=83.2)$ & $m_{\eta'}^0 \ (F^0_{\pi}=93.3)$ \\
\hline
0.5 & 488 & 435 \\
1.0 & 690  & 615  \\
\end{tabular}
\renewcommand{\arraystretch}{1.0}
\end{center}
\end{table}


\begin{references}
\bibitem{1}
   S. Weinberg, Phys. Rev. D 11 (1975) 3583
\bibitem{2}
   G.A. Christos, Phys. Rep. 116 (1984) 251; \\
   W. Marciano, H. Pagels, Phys. Rep. C 36 (1978) 1
\bibitem{3}
   E. Witten, Nucl. Phys. B 156 (1979) 269
\bibitem{4}
   G. Veneziano, Nucl. Phys. B 159 (1979) 213
\bibitem{5}
   G. 't Hooft, Nucl. Phys. B 72 (1974) 461
\bibitem{6}
   E. Witten, Nucl. Phys. B 160 (1979) 57
\bibitem{7}
   M. Teper, Nucl. Phys. B (Proc. Suppl.) 83-84 (2000) 146; \\
   A. Hart, M. Teper, hep-lat/0009008
\bibitem{8}
   P. Gerber, H. Leutwyler, Nucl. Phys. B 321 (1989) 387
\bibitem{9}
   A. Ali Khan et al., (CP-PACS), hep-lat/9909045
\bibitem{10}
   P. van Baal (Ed.). Confinement, Duality and Nonperturbative Aspects of QCD.
NATO ASI Series B: Physics, vol. 368
\bibitem{11}
   K-I. Aoki, O. Miymura, T. Suzuki (Eds.). Non-Perturbative QCD. Structure of the QCD Vacuum,
    Prog. Theor. Phys. Suppl. 131 (1998) 1
\bibitem{12}
   V.N. Gribov, Eur. Phys. Jour. C 10 (1999) 91
\bibitem{13}
   E.V. Shuryak, Phys. Rep. 115 (1984) 151
\bibitem{14}
   V. Gogohia, Gy. Kluge, Phys. Lett. B 477 (2000) 387; \\
   V. Gogohia, Gy. Kluge, Phys. Rev. D 62 (2000) 076008
\bibitem{15}
   R.J. Crewther, Phys. Rev. Lett. 28 (1972) 1421; \\
   M.S. Chanowitz, J. Ellis, Phys. Rev. D 7 (1973) 2490; \\
   J.C. Collins, A. Duncan, S. D. Joglecar, Phys. Rev. D 16 (1977) 438
\bibitem{16}
   V.A. Novikov, M.A. Shifman, A.I. Vainshtein, V.I. Zakharov,
    Nucl. Phys. B 191 (1981) 301
\bibitem{17}
   I. Halperin, A. Zhitnitsky, Nucl. Phys. B 539 (1999) 166
\bibitem{18}
   M.S. Chanowitz, S. Sharpe, Nucl. Phys. B 222 (1983) 211
\bibitem{19}
   J. Hosek, G. Ripka, Z. Phys. A 354 (1996) 177; \\
   Yu.A. Simonov, Phys. Usp. 166 (1996) 337; \\
   H.G. Dosch, Prog. Part. Nucl. Phys. 33 (1994) 121
\bibitem{20}
   T.G. Kovacs, E.T. Tomboulis, Nucl. Phys. B (Proc. Suppl.) 73 (1999) 566; \\ 
   L. Del Debbio, M. Faber, J. Greensite, S. Olenik, Nucl. Phys. B (Proc. Suppl.) 63 (1998) 552; \ A. Ivanov et al., private communication; \\
  P. de Forcrand, M. D'Elia, Phys. Rev. Lett. 82 (1999) 4582
\bibitem{21}
   J.M. Cornwall, Phys. Rev. D 61 (2000) 085012
\bibitem{22}
   T. Schafer, E.V. Shuryak, Rev. Mod. Phys. 70 (1998) 323
\bibitem{23}
   G. 't Hooft, Phys. Rev. Lett. 37 (1976) 8
\bibitem{24}
   V. Gogohia,  H. Toki, Phys. Lett. B 466 (1999) 305; \\
   V. Gogohia,  H. Toki, Phys. Rev. D 61 (2000) 036006
\bibitem{25}
   M.A. Shifman, A.I. Vainshtein, V.I. Zakharov, Nucl. Phys. B 147 (1979) 385  
\bibitem{26}
   E. Meggiolaro, Phys. Rev. D 58 (1998) 085002; \\
   G. Gabadadze, Phys. Rev. D 58 (1998) 094015
\bibitem{27}
   M. Schaden, Phys. Rev. D 58 (1998) 025016
\bibitem{28}
   J.S. Bell, R.A. Bertlmann, Nucl. Phys. B 177 (1981) 218; \\
   R.A. Bertlmann et al., Z. Phys. C 39 (1988) 231
\bibitem{29}
   S. Narison, Phys. Lett. B 387 (1996) 162
\bibitem{30}
   D.I. Diakonov, M.V. Polyakov, C. Weiss, Nucl. Phys. B 461 (1996) 539
\bibitem{31}
   M.-C. Chu, J.M. Grandy, S. Huang, J.W. Negele, Phys. Rev. D 49 (1994) 6039
\bibitem{32}
   T. DeGrand, A. Hasenfratz, T.G. Kovacs, Nucl. Phys. B 505 (1997) 417
\bibitem{33}
   A. Hasenfratz, C. Nieter, Nucl. Phys. B (Proc. Suppl.) 73 (1999) 503
\bibitem{34}
   D.A. Smith, M. J. Teper, Phys. Rev. D 58 (1998) 014505
\bibitem{35}
   P. de Forcrand et al., Nucl. Phys. B (Proc. Suppl.) 63A-C (1998) 549;\\
   P. de Forcrand et al., hep-lat/9802017
\bibitem{36}
   M.G. Perez, T.G. Kovacs, P. van Baal, Phys. Lett. B 472 (2000) 295
\bibitem{37}
   R.C. Brower, D. Chen, J.W. Negele, E.V. Shuryak, Nucl. Phys. B (Proc. Suppl.) 73 (1999) 512
\bibitem{38}
   M.I. Polikarpov, A.I. Veselov, Nucl. Phys. B 297 (1988) 34
\bibitem{39}
   C.G. Callan, Jr., R. Dashen, D.J. Gross, F. Wilczek, A. Zee, Phys. Rev. D 18
(1978) 4684
\bibitem{40}
   A. Hart, M. Teper, Phys. Lett. B 371 (1996) 261
\bibitem{41}
   V. Gogohia, Gy. Kluge,  H. Toki, T. Sakai, Phys. Lett. B 453 (1999) 281
\bibitem{42}
   R. Narayanan, R.L. Singleton, Jr., Nucl. Phys. B (Proc. Suppl.) 63A-C      
(1998) 555
\end{references}
\end{document}